\documentclass[prd,aps,floats]{revtex4}
\usepackage{amsmath}
\usepackage{amssymb}

\numberwithin{equation}{section}

\usepackage{epsfig}

\begin{document}
\def\beq{\begin{equation}}
\def\eeq{\end{equation}}
\def\ber{\begin{eqnarray}}
\def\eer{\end{eqnarray}}
\def\l{\Lambda}
\def\b{{\rm b}}
\def\m{{\rm m}}
\def\om{\Omega_{0\rm m}}
\def\oml{\Omega_l}
\def\omx{\Omega_{\l_{\rm b}}}
\def\eg{8 \pi G}
\def\ppp{{\prime\prime\prime}}
\def\pppp{{\prime\prime\prime\prime}}
\def \lleq {\lower0.9ex\hbox{ $\buildrel < \over \sim$} ~}
\def \ggeq {\lower0.9ex\hbox{ $\buildrel > \over \sim$} ~}
\def\rhoc{\rho_{0 {\rm c}}}
\def\t{\times}
\def\etal{{\it et al.}}
\def\ie{i.e.~}
\def\n{\noindent}

\def\apj{Astroph.~J.~}
\def\mn{Mon.~Not.~Roy.~Ast.~Soc.~}
\def\asta{Astron.~Astrophys.~}
\def\aj{Astron.~J.~}
\def\prl{Phys.~Rev.~Lett.~}
\def\prd{Phys.~Rev.~D~}
\def\nucp{Nucl.~Phys.~}
\def\nat{Nature~}
\def\plb{Phys.~Lett.~B~}
\def\jetpl{JETP ~Lett.~}

\title{A New Universal Local Feature in the Inflationary Perturbation Spectrum}

\author{Minu Joy$^a$, Varun Sahni$^a$ and Alexei A. Starobinsky$^{b,c}$}
\affiliation{$^a$ Inter-University Centre for Astronomy and
Astrophysics, Pune 411~007, India} \affiliation{$^b$ Landau
Institute for Theoretical Physics, 119334 Moscow, Russia}
\affiliation{$^c$ Yukawa Institute for Theoretical Physics, Kyoto
University, Kyoto 606-8502, Japan}

\thispagestyle{empty}

\sloppy

\begin{abstract}
\small{A model is developed in which the inflaton potential
experiences a sudden small change in its second derivative (the
effective mass of the inflaton). An exact treatment demonstrates
that the resulting density perturbation has a quasi-flat power
spectrum with a break in its slope (a step in $n_s$). The step in
the spectral index is modulated by characteristic oscillations and
results in large running of the spectral index localised over a few
e-folds of scales. A field-theoretic model giving rise to such
behaviour of the inflationary potential is based on a fast phase
transition experienced by a second scalar field weakly coupled to
the inflaton. Such a transition is similar to that which terminates
inflation in the hybrid inflationary scenario. This scenario
suggests that the observed running of the spectral index in the WMAP
data may be caused by a fast second order phase transition which
occurred during inflation.}
\end{abstract}

\maketitle

\section{Introduction}
\label{sec:intro} The present decade appears to have ushered in a
golden age for precise cosmological observations. Observations of
the CMB, most recently by the WMAP satellite, combined with
measurements of the matter power spectrum from large scale
structure, weak lensing surveys and Ly-$\alpha$ absorption, permit
parameters of the `standard cosmological model' to be determined to
great accuracy. A consensus appears to be emerging that the late
time behaviour of this model is `close to' LCDM with an
approximately scale invariant primordial spectrum for density
perturbations such as those predicted by the simplest inflationary
models. Indeed, the inflationary paradigm has been remarkably
successful in providing explanations for some well known properties
of the universe including its spatial flatness. It also provides a
mechanism for seeding galaxies by generating an approximately flat
(scale invariant) primordial perturbation spectrum. Both these
predictions of the inflationary scenario have received considerable
observational support from measurements of anisotropies in the
cosmic microwave background (CMB) as detected by WMAP and other CMB
experiments \cite{wmap,bond_boom,archeops}.

However, it is well known that, although primordial fluctuations
spectra expected from inflation are likely to be approximately flat,
or scale-invariant ($n_s(k) \equiv d\ln P_{\cal R}(k)/d\ln k \simeq
1$), exact scale invariance ($n_s=1$) is achievable only for a very
specific class of models \cite{st05}, while a slightly red spectrum
($n_s \lleq 1$) appears to be a generic prediction \cite{pert} of
the simplest viable one-parameter family of inflationary models
including, in particular, $R+R^2$ as well as new and chaotic
inflation \cite{models}. More sophisticated inflationary models
belonging to the slow-roll class may also have $n_s$ slightly
exceeding unity, a notable example being hybrid inflation
\cite{hybrid}. Still for all these models $|n_s(k)-1|\ll 1$, and the
running of the slope $\tilde\alpha(k)\equiv d\,n_s(k)/ d\ln k$ is
also expected to be small : $|\tilde\alpha(k)|\sim |n_s(k)-1|^2\ll
1$ (tilde is used here to avoid confusion with the Bogoliubov
parameter below). Existing CMB and other observational data are just
approaching the level of accuracy necessary to detect deviations
from exact scale invariance and to distinguish between different
inflationary models.

However, while on the one hand the recent WMAP data provide the
first evidence for $n_s$ being close to (but slightly less than)
unity \cite{wmap} (this result is at present inconclusive, see e.g.
\cite{PE06}), on the other, WMAP data \cite{wmap} also suggest a
rather large value of the running $|\tilde\alpha(k)|\sim
|n_s(k)-1|$, and the existence of local spikes like the `Archeops
feature' at $l\sim 40$ (first found in the Archeops data
\cite{archeops} and subsequently confirmed by WMAP \cite{wmap})
which may indicate that inflation is altogether more complex than
the simplest paradigms presented above. Therefore, though it is not
yet clear if these features really exist in the primordial
perturbation spectrum and are not foreground effects or statistical
flukes, it is important to have a list of possible local 'features'
such as bumps, wells, wiggles or spikes, superimposed on an
approximately scale invariant smooth spectrum, which are expected in
more complicated inflationary models. In particular, the large value
of the running, if confirmed, should also be a local feature around
the present Hubble scale since its persistence until the very end of
inflation is incompatible with the requirement that the inflationary
epoch be of sufficient duration (\ie the number of e-folds $N\sim
50$) \cite{EP06}.

Such features in the primordial spectrum of fluctuations are likely to be
measured to great accuracy with next generation CMB satellites such as Planck
\cite{planck}. The precise nature of the primordial spectrum is of
great importance to cosmology not only because it would lead to an in depth
understanding of inflation but also because of its bearing on the values of
the remaining cosmological parameters. (Cosmological parameters
are usually estimated by means of a procedure such as a
{\em Markov Chain Monte Carlo} (MCMC) scheme which assigns probabilities to
cosmological parameters in a multi-dimensional parameter space. Better
knowledge of one set of parameters values  can therefore significantly
influence the probabilities for the remaining.)

A generic way to obtain local deviations from the approximately flat
spectrum in the inflationary scenario is to add additional scalar fields to
the inflaton, either explicitly -- which leads to multiple inflation
\cite{KLS85,KL87,SBB89} or implicitly, through a rapid change in the
effective potential of the inflaton \cite{S92}. In both cases, at
least one of these additional scalar fields should {\em not} be in the
slow-rolling regime, otherwise the spectrum remains approximately flat
\cite{S85}. In this paper we further investigate the second case with
the aim of finding new characteristic features being grafted on
the primordial inflationary spectrum which are less peculiar than
those investigated previously (in view of the observational fact that
there is not too much room for such features).

So, let some scalar field which is weakly coupled to the inflaton
experience a {\em fast} phase transition during inflation, by fast
is meant that its characteristic time is much less than the Hubble
time $H^{-1}(t)$. This phase transition may be induced by the
coupling of the scalar either to the inflaton directly, or to slowly changing
space-time curvature. This requirement usually implies that the
second field be much heavier than the inflaton as well as $H$. Then,
assuming that this transition occurs adiabatically and the second
field is always in the state of local thermodynamical equilibrium
for fixed external values of the inflaton and space-time curvature,
it can be integrated out leaving only some correction to the
inflaton effective potential $V(\varphi)$ that may be non-analytic
(see e.g. \cite{YY06} for detailed derivation).

Following the classification in \cite{S98}, let us consider possible
local non-analytic features of $V(\varphi)$ arising at the point
$\varphi = \varphi_0$ when the fast phase transition discussed above
occurs during inflation. These features can be discontinuities
(jumps) either in $V(\varphi)$ itself or in one of its derivatives.
We denote by $[\,]$ a jump in the relevant quantity, so that $[A]
\equiv A(\varphi_0+0) - A(\varphi_0-0)$. The first three most
interesting cases placed in the order of their decreasing
peculiarity are:

\begin{enumerate}

\item $[V] \neq 0$ -- a step in the effective potential $V(\varphi)$.
This feature is the most peculiar one. Note that it may not arise from
a fast quasi-equilibrium phase transition since that would contradict
energy (more precisely, free energy) conservation. The only way to
obtain such feature is to assume that duration of the phase transition
is comparable to $H^{-1}$, so it is not too fast actually.

Not unexpectedly, this results in a non-universal form of the
corresponding local feature in the adiabatic perturbation spectrum
which depends on detailed dynamics of the phase transition. In the
most generic case, a bump modulated by strong oscillations appears
in the power spectrum \cite{S92,G05}. Significant bumps in the
observable part of the spectrum are excluded though one may
introduce them at very large $k$, close to the end of inflation, to
obtain a significant number of primordial black holes (PBH), see
e.g. \cite{PNN94,GLW96}). If the size of the step is very small, the
bump disappears and only a burst of oscillations remains. The latter
case was studied in \cite{step_pot}. However, the most recent
analysis shows that there is no definite evidence for such a feature
in the observable part of the spectrum \cite{HCMS07}.

\item $[V] = 0,~[V'] \neq 0$. A kink in the potential $V(\varphi)$ leads to
a step in its slope $V'(\varphi)$. The resulting feature in the power
spectrum has a {\em universal} form which does not depend on the details of the fast
phase transition, namely a step with superimposed oscillations which are
not as pronounced as in the previous case \cite{S92}. As pointed in
\cite{S98}, to produce such a discontinuity in $V(\varphi)$, the phase
transition should be of first order.

Though initially such a feature was used to describe an apparent
excess in the rich cluster power spectrum at $k\sim 0.05\,h^{-1}$
Mpc$^{-1}$ \cite{LPS98} ($h$ is the present Hubble constant in terms
of 100 km s$^{-1}$Mpc${^-1}$ and the present value of the scale
factor is taken as unity), it seems now that the only place where
such feature with a significant value of the step in the primordial
spectrum may still exist is in the vicinity of the present Hubble
scale \cite{step_deriv}. Also, as in the previous case, such a
feature may be introduced at scales close to the end of inflation to
produce PBH \cite{BBKP03}.

\item $[V] = [V'] = 0, [V''] \neq 0$ and $|[V'']|\ll H^2$. A sudden
small change in the slope of the potential (a kink) leads to a step
in its second derivative $V''(\varphi)$. This, mildest of all
discontinuities, can be caused by a fast second order phase
transition during inflation \cite{S98}. The last inequality
guarantees that slow-roll inflation continues during and after the
phase transition, in contrast to the case of the hybrid inflation,
or the case $|[V'']|\sim H^2$ considered in \cite{SBB89} where a
second phase transition with the opposite sign of $V''$ had to be
introduced to restore inflation after a short break (that required
much fine tuning, of course). As a result, in contrast to previous
cases, corrections to the primordial power spectrum $P(k)$ arise in
the next to leading order only (at the same order as the
Stewart-Lyth correction \cite{SL93} in case of a smooth inflaton
potential). However, due to the feature in $V(\phi)$, corrections to
$(n_s(k)-1)$ appear to be of the same order as the standard leading
ones while corrections to the running of $n_s(k)$ dominate the
smooth part of $\tilde\alpha(k)$ over a few e-fold interval of
scales around the feature. As in the previous case, they have a {\em
universal} form, too$^1$ {\footnotetext[1] {The second-order phase
transition considered in [\cite{subir}] occurs during a time period
$\delta t \sim H^{-1}$ that results in the temporal breaking of
slow-roll during the transition, in a more peculiar behaviour of the
effective mass in the equation (3.1) for $\xi_k$ and in a step-like
or even bump-like behaviour of the perturbation spectrum similar to
those occurred in the cases 1 and 2.} }.

\end{enumerate}

Most recent observational constraints on the inflaton potential
$V(\phi)$ and the Hubble function $H(\phi)$ obtained using only the
assumption that these functions may be Taylor--expanded up to the
third order in the range of $\phi$ corresponding to the observable
cosmological window of scales ($1-10^4$ Mpc) show that the first two
slow-roll parameters $\epsilon$ and $\delta$ (defined in Sec. 2
below) are really small but the validity of the slow-roll expansion
beyond them is not established, see \cite{LV07} and references to
previous papers therein. The last, third type of peculiarity is just
the one satisfying these conditions. That is why in this paper we
study it in detail and find the universal form of the corresponding
feature. We shall show that a small step in $V''$ leads to a small
step in the primordial spectral index $n_s$ accompanied by
oscillations with a decreasing amplitude.

The plan of the rest of the paper is as follows. In Sec. 2 small
corrections to the background behaviour due to a jump in $V''(\phi)$
are calculated and their contribution to the Sasaki--Mukhanov
equation for scalar perturbations is found. In Sec. 3 an exact
solution for the resulting feature in $P_{\cal R}(k)$ is obtained. A
microscopic model that can produce such a feature in the effective
potential is considered in Sec.4, and the required values of its
parameters are found. Sec. 5 contains conclusions and discussion.

\section{Background cosmology near a feature in the potential}

As discussed in the previous section, we shall examine a model in
which the potential passes through a step-like discontinuity in its
second derivative at time $t_0$ when $\varphi(t_0) = \varphi_0$. In
practice the discontinuity will be smoothed in a small neighborhood
of $\varphi_0$ which we denote by $\varepsilon$. In order to study
the influence of this feature in $V(\varphi)$ on quantities such as
$H(t)$ and $\varphi(t)$, we Taylor expand these quantities around
$t_0$,

\beq
H(t) = H_0 + t \,\dot{H_0} + \frac{t^2}{2!}\,\ddot{H_0}
       + \frac{t^3}{3!} \,\dddot{H}_{\pm}+  \cdots \label{eq:H} \eeq
similarly,
\beq
\varphi(t) = \varphi_0 + t \,\dot{\varphi_0}
             +\frac{t^2}{2!} \,\ddot{\varphi}_{0}
             + \frac{t^3}{3!} \,\dddot{\varphi}_{\pm} + \cdots
\label{eq:varphi}
\eeq
\beq
V^\prime(\varphi) = V^\prime(\varphi_0)
                   + t \,\dot{\varphi_0}\, V^{\prime\prime}_\pm  + \cdots
\label{eq:vprime}
\eeq
for simplicity we have shifted the origin of the time scale to $t_0 = 0$.
The suffix $\pm$ denotes the value of a quantity at $t=t_0\pm\varepsilon \equiv
\pm\varepsilon$.

The equation of motion of the inflaton is
\ber
{\ddot \varphi} &+& 3H{\dot\varphi} + V'(\varphi) = 0~,\nonumber\\
H^2 &=& \frac{8\pi G}{3}\left (\frac{{\dot\varphi}^2}{2} + V(\varphi)\right )~.
\eer
The first of these equations leads to
\beq
\dddot{\varphi}_{\pm} = - 3 H_0 \,\ddot{\varphi}_0
- 3 \dot{H_0} \,\dot{\varphi}_0
- V^{\prime\prime}_{\pm} \, \dot{\varphi}_0~.
\label{eq:3dphi}
\eeq
Next consider the slow roll parameters $\epsilon$, $\delta$,
$\zeta^2$ which are usually defined as
\beq \epsilon = 4\pi G \left(\frac{\dot{\varphi}}{H}\right)^2~, ~~
\delta  = \frac{\ddot{\varphi}}{\dot{\varphi} \, H}~, ~~ \zeta^2 =
\frac{1}{H^2} \left[ \frac{\dddot{\varphi}}{\dot{\varphi}}
          - \frac{\ddot{\varphi}^2}{\dot{\varphi}^2} \right]
\eeq
equivalently
\ber
3 -  \epsilon &=& 8\pi G\frac{V}{H^2}~, \nonumber\\
\delta + 3  &=& -\frac{V^\prime}{H\,\dot{\varphi}}~,\nonumber\\
9 - \zeta^2 &=& 8\pi G\frac{3 V}{H^2} + \frac{3
V^\prime}{H\,\dot{\varphi}}
+\frac{{V^\prime}^2}{H^2\,{\dot{\varphi}}^2} +
\frac{V^{\prime\prime}}{H^2}~, \label{eq:zeta} \eer
where the last equation can be rewritten as \beq \zeta^2 = 3
\epsilon - 3 \delta - \delta^2 - \frac{V^{\prime\prime}}{H^2}~. \eeq
The slow roll condition $\delta \ll 1$ leads to
\beq
\dot{\varphi} \simeq - \frac{V^\prime}{3H}~. \label{eq:slow-roll}
\eeq
However the presence of a feature in the inflaton potential will
result in small corrections to this equation close to $\varphi
\simeq \varphi_0 $. Denoting by $\delta\dot{\varphi}$ the correction
to (\ref{eq:slow-roll}) we find
\ber \delta\dot{\varphi} &=& -\dot{\varphi}_0 \left\{
\frac{\delta_0}{3} +t H_0 \left[\epsilon_0 -\delta_0 +
\frac{\epsilon_0 \, \delta_0}{3} - \frac{V^{\prime\prime}_{\pm}}{3
H^2_0}\right]
\right. \nonumber \\
&& \left.\mbox{} - \frac{t^2 H^2_0}{2} \left[ \left(3\epsilon_0 - 3
\delta_0 -2\epsilon^2_0 - 2 \epsilon_0\,\delta_0 -\frac{2}{3}
\,\epsilon \,\delta_0 \,(\epsilon_0+\delta_0)\right) - \left(3 -2
\epsilon_0 -\delta_0 \right) \frac{V^{\prime\prime}_{\pm}}{3 H^2_0}
+ \sqrt{\frac{2 \epsilon_0}{\eg}} \, \frac{V^{\ppp}_{\pm}}{3 H^2_0}
\right] \right\} \eer
where $\epsilon_0, \delta_0$ are the slow roll parameters evaluated
at $t_0$. The correction to (\ref{eq:slow-roll}) are therefore quite
small.

Next we estimate corrections to the slow-roll parameters, which are found to be
\ber
\epsilon(t) &=& \epsilon_0
+ t H_0 \, \left[2 \epsilon^2_0 + 2 \epsilon_0  \delta_0  \right] \nonumber \\
&&\mbox{} +t^2 H^2_0 \, \epsilon_0 \left[3 \epsilon_0 - 3 \delta_0 -
3 \epsilon^2_0 + 6 \epsilon_0 \delta_0 + \delta^2_0
-\frac{V^{\prime\prime}_{\pm}}{H^2_0} \right]~, \eer
\ber \delta(t) &=&  \delta_0 + t H_0 \, \left[3 \epsilon_0 - 3
\delta_0 + \epsilon_0 \, \delta_0 - \delta_0^2
- \frac{V^{\prime\prime}_{\pm}}{H^2_0}\right] \nonumber \\
&& \mbox{}
+ \frac{t^2 H^2_0}{2}
\left[9 \epsilon_0 + 9 \delta_0 + 6 \epsilon^2_0 + 3 \epsilon_0 \, \delta_0
      + 9 \delta_0^2 + 2 \epsilon_0^2 \,\delta_0  + 2 \delta_0^3 \right. \nonumber \\
&& \left.\mbox{}
 ~~~~~~~~~~~~ + \left(3 -2 \epsilon_0 + 2 \delta_0 \right)
\frac{V^{\prime\prime}_{\pm}}{H^2_0}
- \sqrt{\frac{2 \epsilon_0}{\eg}} \, \frac{V^{\ppp}_{\pm}}{H^2_0}  \right]~,
\eer
\ber
\zeta^2(t) &=&  3 \epsilon_0  - 3 \delta_0 -\delta_0^2 - \frac{V^{\prime\prime}_{\pm}}{H^2_0}\nonumber \\
&& \mbox{} + t H_0 \left[-9 \epsilon_0  + 9 \delta_0 + 6
\epsilon^2_0  + 9 \delta_0^2 - 3 \epsilon_0 \, \delta_0 - 2
\epsilon_0 \, \delta_0^2 + 2 \delta_0^3 + (3 - 2 \epsilon_0 + 2
\delta_0) \frac{V^{\prime\prime}_{\pm}}{H^2_0}
- \sqrt{\frac{2 \epsilon_0}{\eg}} \, \frac{V^{\ppp}_{\pm}}{H^2_0} \right]\nonumber \\
&& \mbox{} +\frac{t^2 H_0^2}{2} \left [27 \epsilon_0  - 27 \delta_0
- 18 \epsilon^2_0 - 63 \delta_0ait^2 + 45 \epsilon_0 \, \delta_0 +18
\epsilon^3_0 + 6 \epsilon^2_0 \, \delta_0 +36 \epsilon_0 \,
\delta^2_0
-6 \epsilon^2_0 \, \delta^2_0 - 36 \delta^3_0  \right. \nonumber \\
&& \left.\mbox{}~~~~~~~~~~~ + 4 \epsilon_0 \, \delta^3_0 - 6
\delta^4_0 -\left(9 - 12\epsilon_0 +  24\delta_0 + 6\epsilon^2_0 +
8\delta^2_0 - 4\epsilon_0 \, \delta^2_0 +
\frac{V^{\ppp}_{\pm}}{H^2_0} \right)\frac{V^{\ppp}_{\pm}}{H^2_0}
\right. \nonumber \\
&& \left.\mbox{}~~~~~~~~~~~ +(3 - 4 \epsilon_0 +\delta_0)
\sqrt{\frac{2 \epsilon_0}{\eg}} \, \frac{V^{\ppp}_{\pm}}{H^2_0} -
\frac{2 \epsilon_0}{\eg} \, \frac{V^{\pppp}_{\pm}}{H^2_0}  \right]~.
\eer
It is well known that the perturbations in the inflaton field and
perturbations in the space-time metric can be reduced to a single
equation either for the gravitational potential $\Phi$ \cite{sasaki83} or the
quantity $\xi$ \cite{mukhanov88}.
We shall use the latter and remind the reader that
$\xi = \delta\varphi - \frac{\dot\varphi}{6H}(\lambda+\mu)$,
where $\lambda$ and $\mu$ describe scalar perturbations of the metric in the
synchronous reference frame  \cite{lifshitz}.
The evolution of the fourier component $\xi_k$ during inflation is described by
the equation
\beq
\ddot{\xi_k} + 3 H \, \dot{\xi_k} + \left(\frac{k^2}{a^2} + m^2_{eff}\right) \,\xi_k = 0
\label{eq:motion}
\eeq
where the effective mass $m^2_{eff}$ is
\beq
m^2_{eff} = \frac{d^2V}{d\varphi^2}
+ 8\,\pi \, G \,\frac{\dot{\varphi}}{H} \, \frac{dV}{d\varphi}
+ H\,\frac{d}{dt}\left(\frac{\dot{H}}{H^2}\right)
\eeq
Using the results obtained earlier in this section and omitting
lengthy intermediate steps, we find the following expression for the
effective mass
\ber \frac{m^2_{eff}(t)}{H_0^2} &=&
\frac{V^{\prime\prime}_{\pm}}{H_0^2} + t H_0 \, \sqrt{\frac{2
\epsilon_0}{\eg}}\,\frac{V^{\ppp}_{\pm}}{H_0^2} \nonumber \\ &&
\mbox{}
- 2 \epsilon_0 \left(3 + \epsilon_0 + 2 \, \delta_0 \right)\, \nonumber \\
&& \mbox{} - 4 \,\epsilon_0 \,t H_0 \left[ \left(3 \,\epsilon_0 +
\epsilon^2_0 + 3\, \epsilon_0 \, \delta_0                  +
\delta^2_0\right) + \frac{V^{\prime\prime}_{\pm}}{H_0^2} \right]
\eer
 The leading term in the right hand side of the above equation
for $t\to 0~(\phi\to\phi_0)$ which is of the first order in
$\epsilon,\delta$ is $V^{\prime\prime}_\pm/H_0^2 -6\epsilon_0$. The
perturbation equation, Eq.~(\ref{eq:motion}), with $m^2_{-} =
V^{\prime\prime}_- -6\epsilon_0$ when $t<0$ and $m^2_{+} =
V^{\prime\prime}_+ - 6\epsilon_0$ when $t>0$ (so that $[m^2]=[V'']$
since $\epsilon_0$ is continuous at $t=0$), therefore provides an
excellent approximation to the dynamics. This is the main result of
this section.
\section{Perturbation spectrum and spectral index }
\label{sec:bogolubov}
As demonstrated in the previous section, the jump in the effective
mass is equal to the jump in the second derivative of the inflaton
potential: $[\Delta m_{\rm eff}^2]=[\Delta V'']$. Next consider the
motion of the inflaton as it rolls down its potential.
If the feature is crossed by the inflaton at $t_0$, then at $t \ll
t_0$ as well as $t \gg t_0$ the slow roll condition $|V''| \ll 24\pi
GV$ remains valid, which permits us to solve (\ref{eq:motion}) as if
the effective mass were constant. In terms of the conformal time
coordinate $\eta = \int dt/a(t)$ equation (\ref{eq:motion}) acquires
the form
\beq \xi_k'' + 2\frac{a'}{a}\xi_k' + \left (k^2 + m_{\rm
eff}^2a^2\right )\xi_k = 0~, \eeq
where the derivatives are with respect to $\eta$. The transformation
$\xi_k = \chi_k/a$ results in an oscillator-type equation in which
the frequency is time dependent
\beq \chi'' + \left (k^2 + m_{\rm eff}^2a^2 - \frac{a''}{a}\right )
\chi = 0~, \label{eq:perturb1} \eeq
where we have dropped the suffix $k$ in $\chi_k$ for simplicity. In
passing note that equation (\ref{eq:perturb1}) is equivalent to
\beq \label{em2}\chi'' + \left (k^2 - \frac{z''}{z}\right )\chi =
0~, ~~{\rm where} ~~ z = \frac{a{\dot\phi}}{H}~, \eeq
which implies that on large scales ($k^2 \ll z''/z),~ \chi/z \to$
constant.
In the following discussion we assume that the discontinuity in the
second derivative of the potential is reached by the field
$\varphi(t)$ at the time $t=t_0$ ($\eta = \eta_0$). The normalized
solution to (\ref{eq:perturb1}) corresponding to the adiabatic
vacuum at early times ($t \ll t_0$, equivalently $\eta \ll \eta_0$)
is
\beq \chi_{\rm in}(\eta) =
\frac{\sqrt{\pi\eta}}{2}H_{\mu_1}^{(2)}(k\eta)~, \eeq
where $H_\mu^{(2)}(k\eta)$ is the Hankel function and ${\mu_1} =
\frac{3}{2} - \frac{V''_-}{3H_0^2} + 3 \epsilon_0 $, where $V''
\equiv \frac{d^2V}{d\varphi^2}$. (We assume that the expansion of
the universe is quasi-exponential.) The behaviour of perturbations
{\em after} the feature is crossed ($t \gg t_0$, $\eta \gg \eta_0$)
will be described by a superposition of positive and negative
frequency solutions of (\ref{eq:perturb1}), namely
\beq \chi_{\rm out}(\eta) = \frac{\sqrt{\pi\eta}}{2}\left (\alpha
H_{\mu_2}^{(2)}(k\eta) + \beta H_{\mu_2}^{(1)}(k\eta) \right )~,
\label{eq:bog1} \eeq
${\mu_2} = \frac{3}{2} - \frac{V''_+}{3H_0^2} + 3 \epsilon_0$ and
 $\alpha,\beta$ are the Bogoliubov
coefficients. Note the following relationship between $\mu$ and the
scalar spectral index $n$
\beq \mu_{1,2} = 2 - \frac{n_{1,2}}{2}~, \label{eq:index} \eeq
where $n_1 (n_2)$ is the spectral index in the `in' (`out') region.

\begin{figure}[tbh!]
\centerline{ \psfig{figure=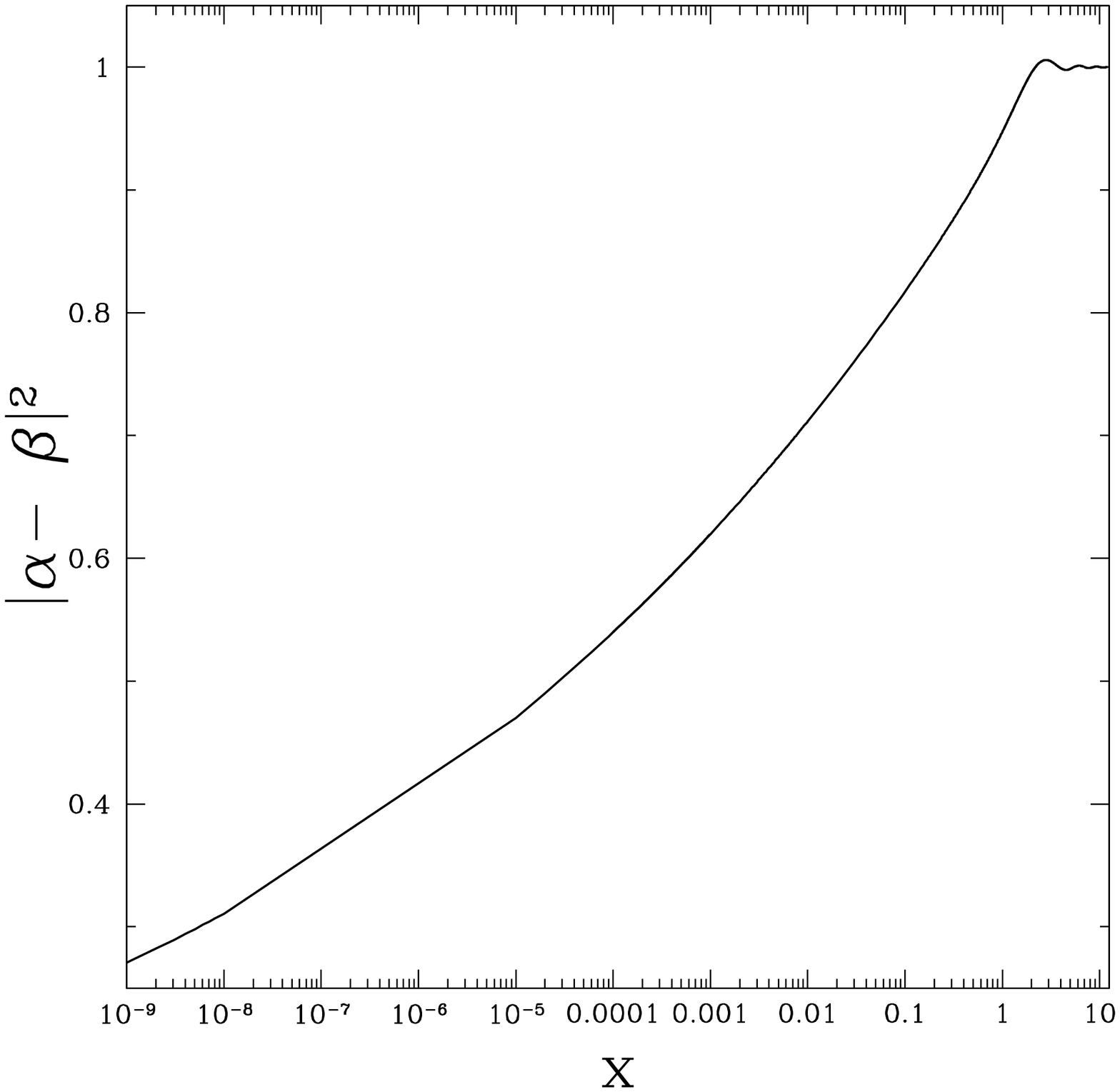,width=0.5\textwidth,angle=0}
\psfig{figure=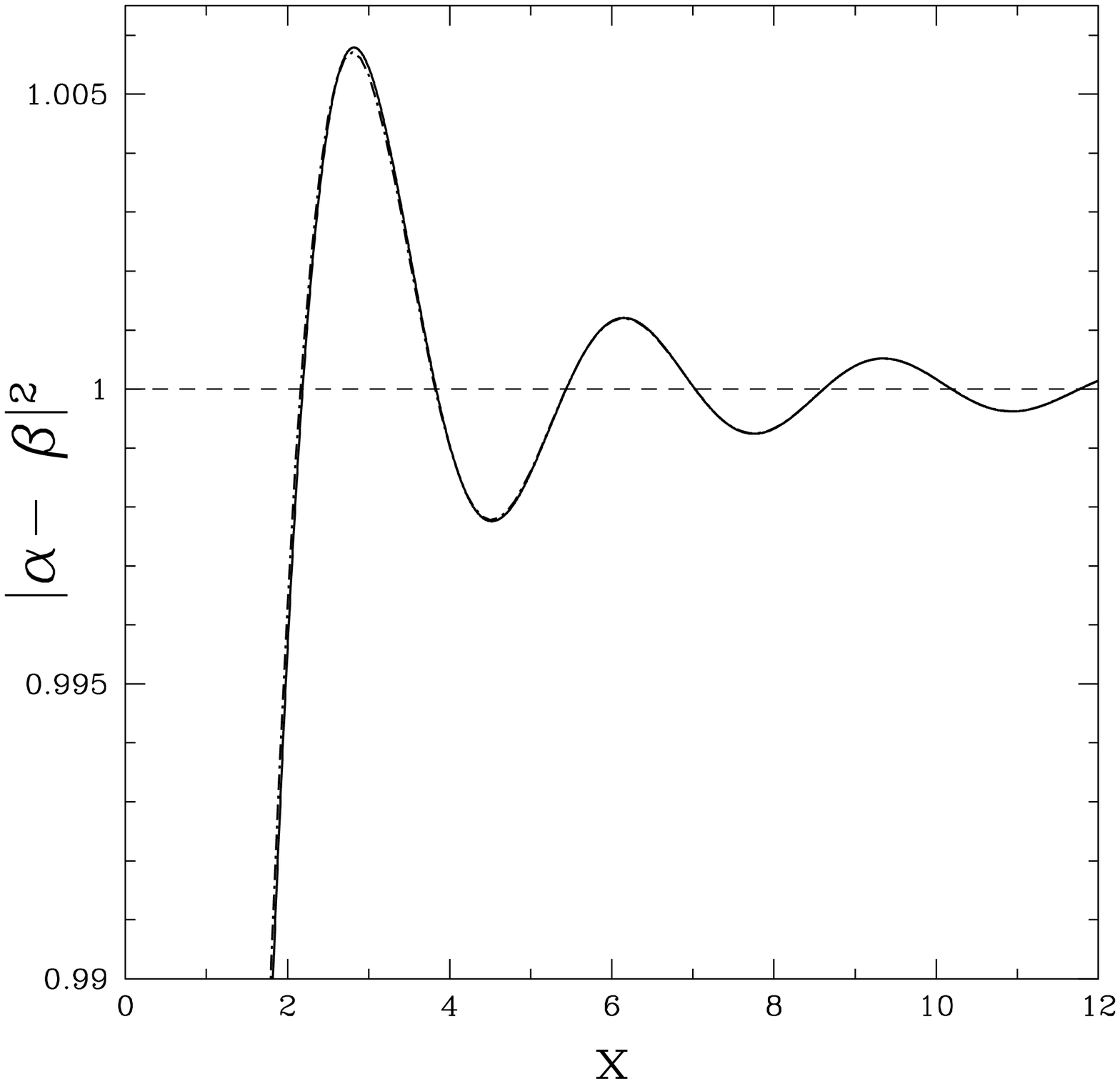,width=0.5\textwidth,angle=0} }
\caption{\small The transfer function
$\left\vert\alpha-\beta\right\vert^2$ is shown as a
function of $x=k/k_0$. 
The exact expression for $\left\vert\alpha-\beta\right\vert^2$ given
in (\ref{eq:bog3}) is represented by the solid line in the left and
right panels while the asymptotic expression (\ref{eq:asymp1}) is
shown dot-dashed in the right panel which shows the oscillations in
the transfer function in greater detail. Note that the asymptotic
expression provides an excellent approximation to the results for
$x=k/k_0 \ggeq 2$. The feature associated with the step in
$V''(\phi)$ occurs at $x \sim 1$. The relevant values of the
parameters are $\mu_1 = 1.49, \,\mu_2 = 1.52$. \label{fig:bog}}
\end{figure}

\begin{figure}[tbh!]
\centerline{\psfig{figure=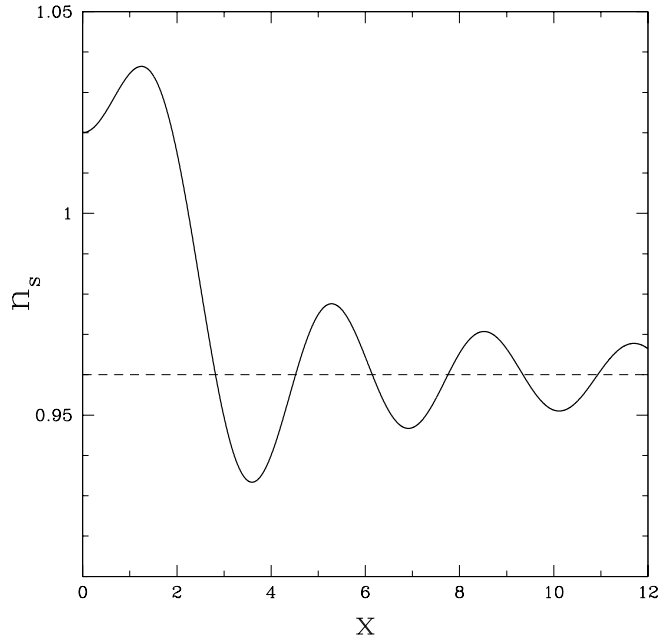,width=0.5\textwidth,angle=0} }
\caption{\small The primordial spectral index $n_s$ is shown as a
function of $x = k/k_0$ for an inflationary model in which the
potential has a sudden change in its second derivative. Such a
discontinuity in $V''$ leads to step in $n_s$ at $x \sim 1$ which is
followed by oscillations with decreasing amplitude described by
(\ref{eq:index2}). The parameters of our model are $\mu_1 = 1.49,\,
\mu_2 = 1.52$ which correspond to $n_1 = 1.02$, $n_2 = 0.96$. }
\label{fig:ns}
\end{figure}

In order to determine $\alpha$ and $\beta$ we
match $\chi_{\rm in} = \chi_{\rm out}$ and ${\chi'}_{\rm in} =
{\chi'}_{\rm out}$ at $\eta=\eta_0$ to obtain
\ber
\alpha-\beta &=& -\frac{i\pi \Delta}{2}H_{\mu_1}^{(2)}(k\eta_0)J_{\mu_2}(k\eta_0)
-\frac{i\pi k\eta_0}{2}\left \lbrack H_{\mu_1+1}^{(2)}(k\eta_0)J_{\mu_2}(k\eta_0)
-H_{\mu_1}^{(2)}(k\eta_0)J_{\mu_2+1}(k\eta_0)\right\rbrack~,\\
\nonumber\\
\alpha+\beta &=& \frac{\pi \Delta}{2}H_{\mu_1}^{(2)}(k\eta_0)Y_{\mu_2}(k\eta_0)
+ \frac{\pi k\eta_0}{2}\left\lbrack H_{\mu_1+1}^{(2)}(k\eta_0)Y_{\mu_2}(k\eta_0)
-H_{\mu_1}^{(2)}(k\eta_0)Y_{\mu_2+1}(k\eta_0)\right\rbrack~,\\
\nonumber\\
|\alpha|^2 - &|\beta|^2& = 1~, \label{eq:bog2} \eer
where $\Delta=\mu_2-\mu_1$. The quantity $|\beta|^2$ corresponds to
the number of scalar particle pairs carrying momenta ${\vec k}$,
$-{\vec k}$ created due to the rapid variation in $V''$ as the
inflaton $\varphi$ crosses the feature at $\eta = \eta_0$. However,
the quantity of interest is related to the late time behaviour of
$\xi_k(t\to \infty)$, namely $\xi_k(\eta\to 0) \equiv
\frac{\chi_{\rm out}(\eta\to 0)}{a} 
\propto (\alpha-\beta)$, which contributes to the growing mode of
scalar adiabatic perturbations. The corresponding power spectrum for
the curvature perturbations is simply $P_{\cal R}(k) =
\left(\frac{H}{\dot{\phi}} \right)^2 \big\vert\xi_k(\eta_{\to
0})\big\vert^2$. It is important to note that
\beq P_{\cal R}(k) \propto {\mathcal{P_R}}_0(k)\times
|\alpha-\beta|^2 \label{eq:power_spectrum} \eeq
where $P_{\cal R}(k) \propto k^{n_s-1}$ and $\mathcal{P_R}_0(k)$ is
the power specrum of the background model on which the transfer
function $|\alpha-\beta|^2$, describing the feature, has been
overlayed. In our case $\mathcal{P_R}_0(k) \propto k^{n_2-1}$ where
$n_2$ is the spectral index in the `out' region; see
(\ref{eq:index}). Now, substituting $H_{\mu}^{(2)}(z) \simeq
-H_{\mu}^{(1)}(z) \simeq \frac{i}{\pi} \Gamma(\mu)
\left(\frac{z}{2}\right)^{-\mu} $ as $ z \rightarrow 0 $ , into
(\ref{eq:bog1}) we get,
\beq \label{sl}
\mathcal{P_R}_0(k) = \frac{2^{2\mu_2 - 3}}{\pi^3}
\Gamma^2(\mu_2) (1 - \epsilon)^{2\mu_2 - 1}  \,
\left(\frac{H^2}{\big\vert\dot{\phi}\big\vert}\right)^2\, \Bigg
\vert_{\,aH=k}\, ,\eeq when $ \eta \to 0 $.
It is clear that the above expression is in agreement with Eq.~(60)
of \cite{SL93} (Stewart-Lyth correction). The transfer function
$|\alpha - \beta|^2$ differs from unity by terms of order
$\mu_2-\mu_1$ only.
Indeed, from (\ref{eq:bog2}) one readily finds, for the transfer function
\ber \frac{4}{\pi^2}\left \vert\alpha-\beta\right\vert^2
&=& \Delta^2 J_{\mu_2}^2\left (Y_{\mu_1}^2+J_{\mu_1}^2\right )  \nonumber\\
\nonumber\\
&+&(k\eta_0)^2\left\lbrace J_{\mu_2}^2\left
(Y_{\mu_1+1}^2+J_{\mu_1+1}^2\right ) + J_{\mu_2+1}^2\left
(Y_{\mu_1}^2+J_{\mu_1}^2\right ) - 2J_{\mu_2}J_{\mu_2+1}\left
(Y_{\mu_1}Y_{\mu_1+1}+J_{\mu_1}J_{\mu_1+1}\right )
\right\rbrace \nonumber\\
\nonumber\\
&+& 2\Delta~ (k\eta_0) J_{\mu_2}\left\lbrace J_{\mu_2}\left
(Y_{\mu_1}Y_{\mu_1+1}+J_{\mu_1}J_{\mu_1+1}\right ) -
J_{\mu_2+1}\left (Y_{\mu_1}^2+J_{\mu_1}^2\right )\right\rbrace
 \label{eq:bog3} \eer
where the Bessel functions are evaluated at $x=k/k_0$, $k_0$ is the
mode just leaving the Hubble radius at the time of the transition in
$V''$. The functional dependence of $\left
\vert\alpha-\beta\right\vert^2$ on $x$ is shown in figure
\ref{fig:bog} for model parameters $\mu_1 = 1.49, \mu_2 = 1.52$. The
expression for $\left \vert\alpha-\beta\right\vert^2$ in
(\ref{eq:bog3}) has the following useful asymptotic forms:
\begin{enumerate}
\item For $x = k/k_0 \gg 1$
\beq \left \vert\alpha-\beta\right\vert^2 \simeq
1 + \frac{3 \, (\mu_2-\mu_1)}{2x^2}\left \lbrack\cos{(2x)}
- \frac{\sin(2x)}{x}\right \rbrack ~, \label{eq:asymp1} \eeq
\item For $ 0.01 \lleq x \lleq 1 $
\beq \alpha - \beta  \simeq  1 +  (\mu_2-\mu_1)\left[\log{\left
(\frac{x}{2}\right )} + \frac{1}{3} - \psi(5/2)\right]~,
\label{eq:asymp2} \eeq
here $\psi(z)$ is the digamma function, $\psi(z) =
\frac{d}{dz}\log{\Gamma(z)}$, $\psi(5/2) = \frac{8}{3} -2\log{2} -
\gamma$, where $\gamma \simeq 0.5772$ is Euler's constant. And,
%
\item For $x \ll 1$
\beq \left\vert\alpha-\beta\right\vert^2 \simeq
\left (\frac{x}{2}\right )^{2(\mu_2-\mu_1)}
~. \label{eq:asymp3} \eeq
\end{enumerate}
The effective spectral index $n_s(k) \equiv d\ln P_{\cal R}(k)/d\ln
k$ can be determined from the power spectrum,
(\ref{eq:power_spectrum}) as follows
\beq
n_s(k) - 1 = n_2(k) - 1 +
\frac{d\log{\left(\left\vert\alpha-\beta\right\vert^2\right )}}{d\log{k}}~,
\eeq
where the asymptotic forms for $\left\vert\alpha-\beta\right\vert^2$ in
(\ref{eq:asymp1}), (\ref{eq:asymp2}) and (\ref{eq:asymp3})
lead to the following useful approximations
%
\begin{enumerate}
\item For $x = k/k_0 \gg 1$
\beq
\frac{d\log{\left (\left\vert\alpha-\beta\right\vert^2\right
)}}{d\log{k}}
\simeq{-\frac{3\,(\mu_2 - \mu_1)}{x}}\left[\sin{(2x)} +
\frac{2\cos{(2x)}}{x} \right] ~, \label{eq:index1} \eeq
From (\ref{eq:index}), (\ref{sl}) and (\ref{eq:index1}) we find, for
the spectral index
\beq n_s(k) \simeq 1 - 4 \epsilon_0 - 2 \delta_0 - {3\,(\mu_2 -
\mu_1)}~\frac{\sin{(2x)}}{x}~. \label{eq:index2} \eeq
\item For $x \ll 1$
\beq
\frac{d\log{\left (\left\vert\alpha-\beta\right\vert^2\right
)}}{d\log{k}}\simeq 2\left (\mu_2-\mu_1\right )~, \eeq so that \beq
n_s \simeq n_2 + 2\left (\mu_2-\mu_1\right )=n_1~. \eeq
\end{enumerate}
The preceeding discussion has been quite general. In order
to explore our model in more detail we need to give values to its
parameters. Accordingly we set $\mu_1 = 1.49, \mu_2 = 1.52$ which correspond
to $n_1 = 1.02$, $n_2 = 0.96$ -- see eqn. (\ref{eq:index}).
The functional form of $n_s(k)$ for this model is shown in figure
\ref{fig:ns}. Our choice of the spectral indices is largely for a
descriptive purpose, other values can easily be accomodated by the
model. In the final analysis, a judicious choice of $n_1$ and $n_2$
must stem from a comparison of this model with observations that
lies outside the scope of the present paper.
From figure \ref{fig:ns} we see that a discontinuity in the second
derivative of the inflaton potential (a step) leads to a step in the
spectral index, which in our case drops from $n_s = 1.02$ at $k/k_0
\ll 1$ to $n_s = 0.96$ at $k/k_0 \gg 1$. The step in $n_s$ is
accompanied by `ringing' -- slowly decreasing oscillations in $n_s$
about the mean (asymptotic) value of $n_s = 0.96$. From
(\ref{eq:index2}) we also find the following expression for the
running of the spectral index
\beq \tilde\alpha\equiv\frac{dn_s}{d\log{k}} = \frac{dn_2}{d\log{k}}
- 6\,(\mu_2 - \mu_1)\cos{(2x)}~, ~~x = \frac{k}{k_0} \gg 1~.
\label{eq:running} \eeq
From figure \ref{fig:ns} we find that the
running $\tilde\alpha$ is of the order of $n_s-1$ at $x \sim 1$. For $x \gg 1$
equation (\ref{eq:running}) informs us that $\tilde\alpha$ has two components,
of which the second is of order $n_s-1$ but oscillates, while the first
is smooth and of order $(n_s-1)^2$,
as is usually the case in the slow-roll regime.
The following simple relationship between the CMB
multipole $\ell$ and the comoving wavenumber $k$  helps relate the feature
in our potential with a corresponding angular scale $\theta$
\beq
\ell \sim \theta^{-1} \simeq k(\eta_0 - \eta_{\rm ls})~,
\eeq
where
\beq
\eta_0 - \eta_{\rm ls} = \frac{c}{H_0}\int_0^{z_{\rm ls}}{\frac{dz}{h(z)}}~,
\eeq
and $z_{\rm ls}$ is the redshift of the last scattering surface.
This leads to
\beq
k \simeq \ell \times 10^{-4} h ~{\rm Mpc}^{-1}~,
\eeq
in a spatially flat LCDM cosmology with $\Omega_m \simeq 0.3$ and $\Omega_{\Lambda}
\simeq 0.7$.
\section{Inflationary model with a step-like discontinuity in the evolution of the
effective mass}
\label{sec:hybrid}
As an example of a microscopic field-theoretic model which can give
rise to the feature in the inflaton potential discussed in this
paper, let us consider the standard model used in many occasions, in
particular, in the hybrid inflationary scenario \cite{hybrid}:
\beq V(\psi,\phi) = \frac{1}{4\lambda}\left (M^2 -
\lambda\psi^2\right )^2 + \frac{1}{2}m^2\phi^2 +
\frac{g^2}{2}\phi^2\psi^2~. \label{eq:hybrid} \eeq
From (\ref{eq:hybrid}) we find that, near $\psi = 0$, the effective
mass of the field $\psi$ is given by
\beq m_\psi^2 \equiv \frac{d^2V}{d\psi^2} = g^2\phi^2 -M^2~, \eeq
so that $m_\psi^2  > 0$ if $\phi > \phi_c$ and $m_\psi^2  < 0$ if
$\phi < \phi_c$, where $\phi_c = M/g$ is the critical value of the
field $\phi$ at which the curvature of the potential $V(\psi,\phi)$
along the $\psi$ direction vanishes. The change in the sign of
$m_\psi^2$ is a crucial ingredient of this model: $m_\psi^2  > 0$
ensures that at early times the field $\psi$ rolls towards $\psi =
0$, whereas $m_\psi^2  < 0$ at late times, destabilizes the $\psi =
0$ configuration resulting in a rapid cascade (waterfall) of $\psi$
towards the minimum of its potential. Thus, just before the (weakly
second order) phase transition, $\phi > M/g$, $\psi = 0$ so that
\beq V(\phi)= \frac{M^4}{4\lambda} +\frac{m^2\phi^2}{2}~,
\label{eq:pot_before} \eeq
and ${\partial^2 V}/{\partial \phi^2} = m^2$. Introducing the
parameter $\kappa = 2\lambda m^2/g^2M^2$, we write $V(\phi)$ at the
instant of transition as
\beq V(\phi_c) = \frac{M^4}{4\lambda}(1+\kappa)~. \eeq
A large value $\kappa > 1$ implies that the correction from the
$m^2\phi^2$ term to the vacuum energy density $V(0,0) =
M^4/4\lambda$ is significant, while the opposite is true for $\kappa
< 1$. As we shall discover, $\kappa \lleq 1$ will be more relevant
to the scenario which we are considering.
It is easy to see that prior to the transition the slow-roll
condition $\left \vert V''\right \vert/H^2  \ll 1$ implies
\beq M^2 \gg \sqrt{\frac{3 \lambda}{2 \pi}} \, \frac{m
\,m_P}{(1+\kappa)^{1/2}}~, \label{k1} \eeq
where $m^2_P = G^{-1} $.  Soon after the transition, $\phi < M/g$,
$\psi^2=(M^2-g^2\phi^2)/\lambda$ and
\beq V(\phi)=\frac{1}{2}(m^2+\frac{g^2M^2}{\lambda})\phi^2
-\frac{g^4\phi^4}{\lambda}~. \label{eq:pot_after} \eeq
The requirement that slow-roll remain valid immediately after the
transition gives
\beq M\gg gm_P~. \label{k2} \eeq
The product of (\ref{k1}) and (\ref{k2}) results in the following
constraint
\beq M^3 \gg
\sqrt{\frac{3\lambda}{2\pi}}~\frac{gmm_P^2}{\sqrt{1+\kappa}}~.
\label{k1k2} \eeq
Unlike the field $\phi$ which slowly rolls down the potential
$V(\phi,\psi)$, the motion of $\psi$ is rapid and the condition
$\frac{\left\vert \partial^2V/\partial\psi^2\right\vert}{H^2} \gg 1$
is valid all time apart from a very small interval $ \Delta t\ll
H^{-1} $ around the transition if
\beq M^3 \ll \frac{\lambda m \,m_P^2}{1+\kappa}~. \label{k3} \eeq
It is easy to see that (\ref{k1k2}) and (\ref{k3}) imply $g^2 \ll
\lambda$ in our model, i.e. self-coupling of the $\psi$ field should
be much more than its coupling to the inflaton $\phi$.
The value of the spectral index before ($n_1$) and after ($n_2$) the
transition can be determined quite simply, by applying the well
known formula
\beq n-1 = -\frac{3\,m_P^2}{8\pi}\left (\frac{V'}{V}\right )^2 +
\frac{m_P^2}{4\pi }\left (\frac{V''}{V}\right )~, \eeq
to (\ref{eq:pot_before}) and (\ref{eq:pot_after}). Consequently
\ber n_1 - 1 &=& \frac{1}{2\pi}\left (\frac{g m_P}{M}\right
)^2\frac{\kappa~(1-2\kappa)}
{(1+\kappa)^2} ~,\label{eq:n1}\\
n_2-1 &=& -\frac{1}{2\pi}\left (\frac{g m_P}{M}\right)^2
\frac{4+3\kappa+2\kappa^2}{(1+\kappa)^2} ~. \label{eq:n2} \eer
From (\ref{eq:n1}) we find that the inflationary spectrum on large
scales has a red (blue) tilt if $\kappa > 1/2$ ($\kappa < 1/2$);
$\kappa = 1/2$ results in precise scale-invariance for the initial
spectrum: $n_1 = 1$. The total change in the spectral index during
the course of the transition is given by
\beq \Delta n \equiv n_1-n_2 = \frac{2}{\pi (1+\kappa)}\left
(\frac{gm_P}{M}\right)^2~. \label{eq:delta_n} \eeq
Clearly, in order to make contact with observations the value of
$\kappa$ must not be too large since otherwise $n_1 \simeq n_2$, and
it will be difficult to test the predictions of this model
rigorously.
\begin{figure}[tbh!]
\centerline{
\psfig{figure=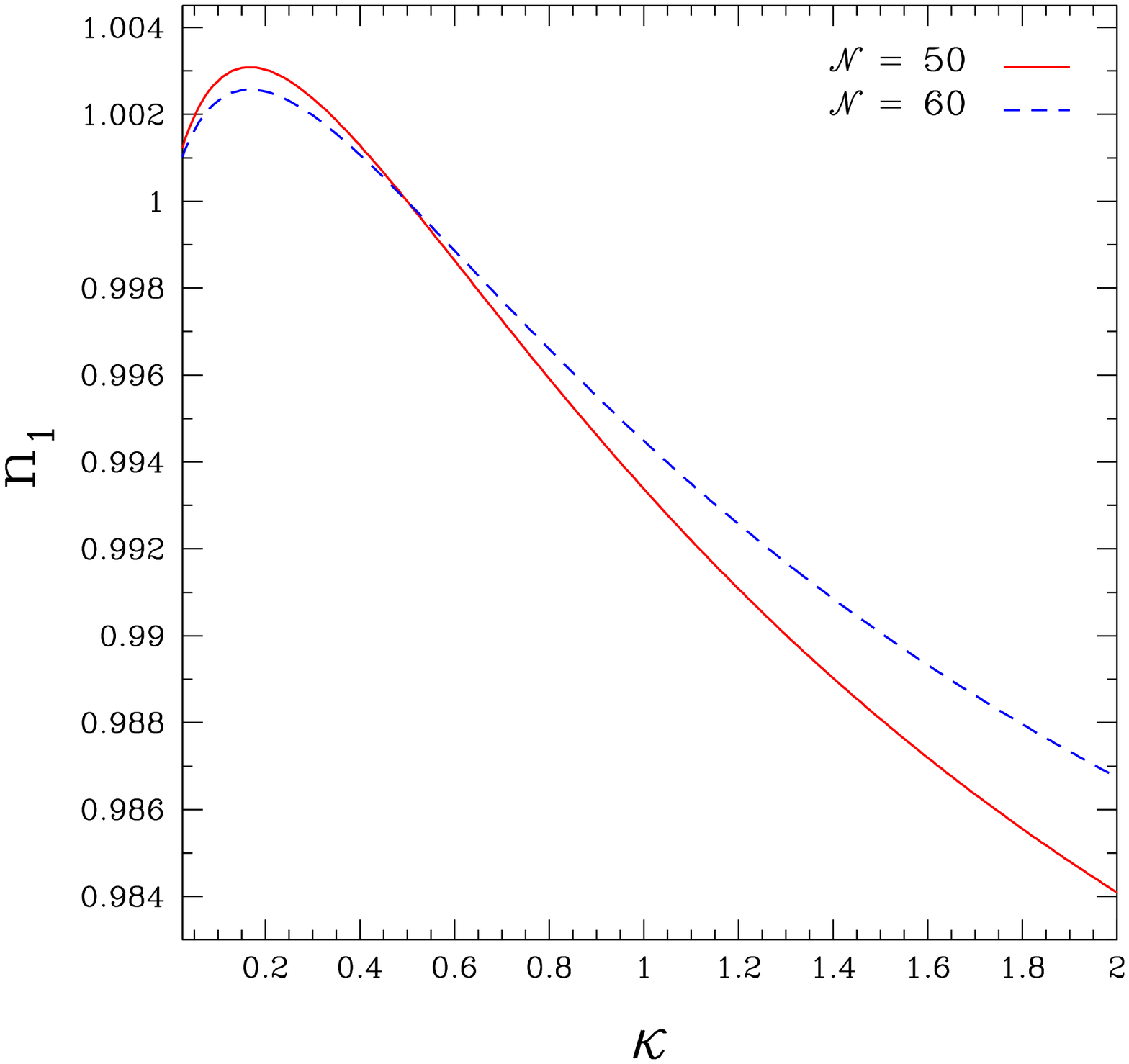,width=0.47\textwidth,angle=0}
\psfig{figure=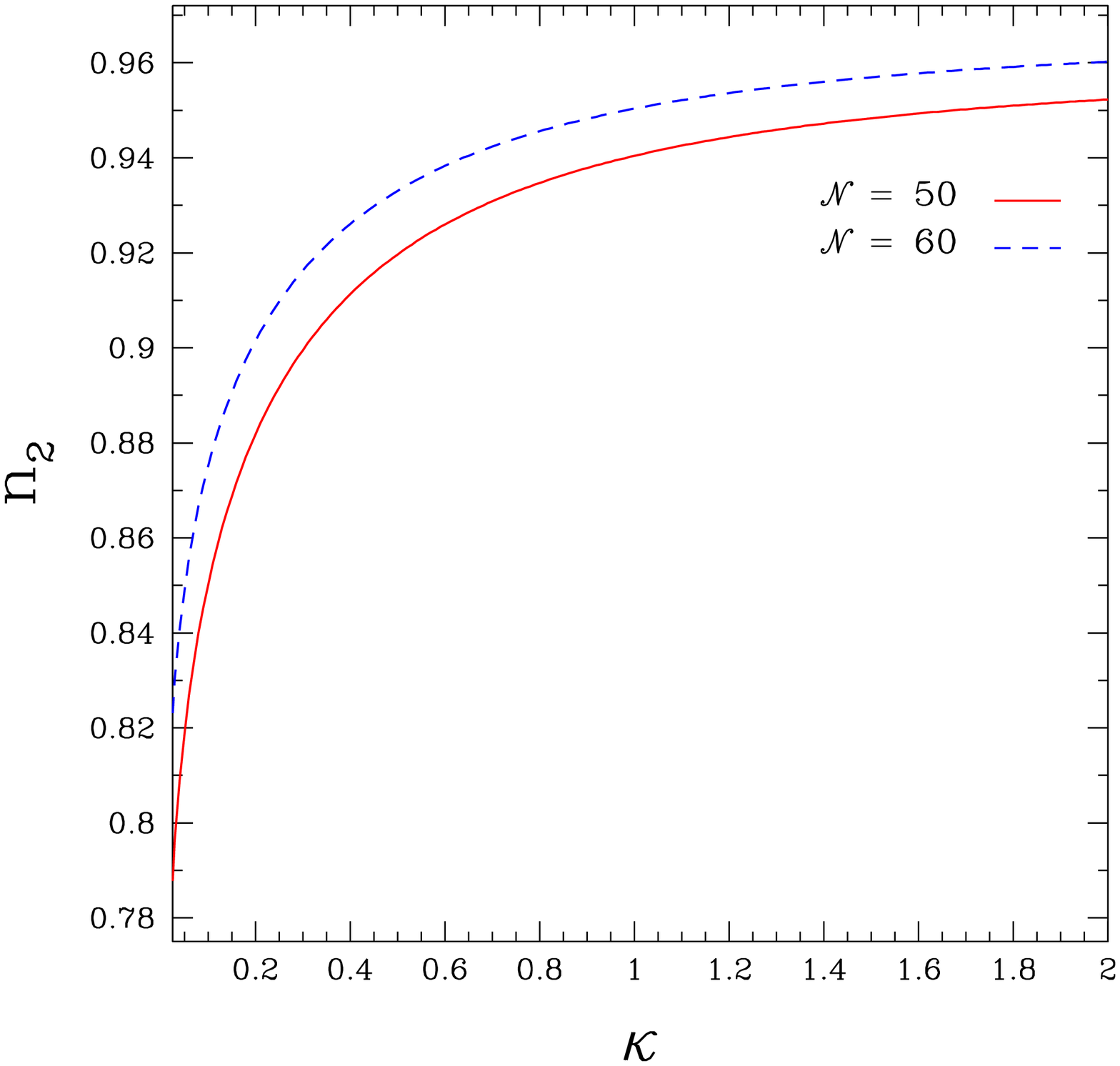,width=0.47\textwidth,angle=0} }
\caption{\label{fig:n1-alpha} The spectral index just before ($n_1$)
the phase transition in hybrid inflation and immediately after it
($n_2$), is shown as a function of the parameter $\kappa = 2\lambda
m^2/g^2M^2$ in the left and right hand panel of this figure. The red
(solid) line corresponds to 50 e-folds of inflationary expansion
occuring after the phase transition in hybrid inflation, while the
dashed (blue) line corresponds to 60 e-folds.}
\end{figure}
\begin{figure}[tbh!]
\centerline{ \psfig{figure=
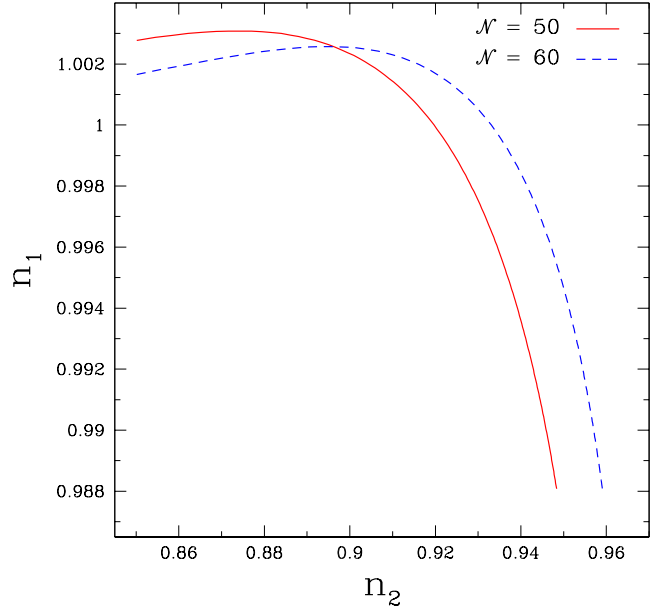,width=0.5\textwidth,angle=0} }
\caption{\label{fig:n1n2} Spectral indices for perturbations
generated just before ($n_1$) and immediately after ($n_2$) the
phase transition in hybrid inflation are shown. Note that distinct
values of the pair $\lbrace n_1,n_2\rbrace$ correspond to different
values of the parameter $\kappa$, as shown in figure
\ref{fig:n1-alpha}. Results are shown for two possible number of
e-folds after the phase transiton: ${\cal N} = 50$ (red, solid), and
${\cal N} = 60$ (blue, dashed). }
\end{figure}
From (\ref{eq:n1}) and (\ref{eq:n2}) we find that $n_1$ and $n_2$
are completely determined if we know the value of $\kappa$ ($\equiv
2\lambda m^2/g^2M^2$) as well as $g m_P/M$. $\kappa$ and $g m_P/M$
are also related to the number of inflationary e-folds which take
place after the phase transition has occurred,
\beq {\cal N} = \pi \left ( \frac{M}{gm_P}\right )^2\left [1 + \left
(1+\frac{\kappa}{2}\right ) \log \frac{2+\kappa}{\kappa}\right]~.
\label{eq:N} \eeq
The exact value of ${\cal N}$ can be chosen as one of free
parameters of the model. It is uniquely defined by the preferred
location $k_0$ of the feature and the length of the reheating period
after inflation. Below we use ${\cal N}=60$ as an estimate which
roughly corresponds to $k_0$ being of the order of the inverse
present Hubble scale, leaving the determination of the most probable
location of $k_0$ needed to explain running in the WMAP data for
future work. As follows from Fig. 2 the transition, from the slope
value $n_s=n_1$ to $n_s=n_2$, mainly occurs in the range $k\approx
(1-3)k_0$ (neglecting small oscillations for higher values of $k$).

 We plot the behaviour of  the spectral index as a function of
$\kappa$ for two different values of ${\cal N}$ in figures
\ref{fig:n1-alpha}, \ref{fig:n1n2}.
Values of parameters ($M, m, g$) in our model can be very simply
estimated by comparing the inflationary curvature fluctuation on
large scales with the observed CMB fluctuation measured by COBE or
WMAP. The perturbation spectrum can be approximated as
\beq P_{\cal R}(k) =
\frac{128\pi}{3m_P^6}\frac{V^3}{V'^2}=\frac{8\pi}{3}
\frac{g^4}{\lambda} \left(\frac{M}{gm_P}\right)^6
\frac{(1+\kappa)^3}{\kappa^2}~, \eeq
which is related to the dimensionless density contrast $\delta_H(k)$
via
\beq \delta^2_H(k) = \frac{4}{25}\left
(\frac{g(\Omega_m)}{\Omega_m}\right )^2 P_{\cal R}(k)~, \eeq where
\cite{carroll92} \beq g(\Omega_m) = \frac{5}{2}\Omega_m\left
(\frac{1}{70} + \frac{209\Omega_m - \Omega_m^2}{140} +
\Omega_m^{4/7}\right )^{-1}~, \eeq
and the 4 year COBE data implies, for an LCDM Universe
\cite{bunn96,ll00},
\beq \delta_H = 1.91\times 10^{-5}~ \frac{\exp{[1.01
(1-n)]}}{\sqrt{1+f(\Omega_m) r}} \Omega_m^{-0.8-0.05\log{\Omega_m}}
\times\left \lbrace 1-0.18(1-n)\Omega_\Lambda-0.03 r
\Omega_\Lambda\right \rbrace~, \eeq
where the ratio of tensor to scalar power spectrum denoted by $r
\simeq 16 \, \epsilon_0$\,, $f(\Omega_m) = 0.75 -
0.13\Omega_\Lambda^2$.
Substituting, for $V(\phi)$ from (\ref{eq:pot_before}) we obtain,
after setting ${\cal N} = 60$ in (\ref{eq:N}), and $\lambda = 0.1$,
for a spatially flat LCDM universe with $\Omega_m = 0.26$ and
$\Omega_\Lambda = 0.74$, parameter values $\kappa,g,M,m$ given in
table 1.
\begin{table}[ht]
\caption{\footnotesize Typical parameter values for the hybrid inflationary model
(\ref{eq:hybrid}) with a step in the spectral index.
}
\medskip

\begin{center}
\begin{tabular}{|c|c|c|c|}
\hline \hline
$\kappa$    &  $g$                     &  $M/m_p$                  & $m/m_p$  \\
\hline
   1        &  3 $\t 10^{-4}$    &  8 $\t 10^{-4}$       & 5.3 $\t 10^{-7}$ \\
   0.5      &  2.9 $\t 10^{-4}$  &  7.2 $\t 10^{-4}$     & 3.3 $\t 10^{-7}$ \\
   0.25     &  2.6 $\t 10^{-4}$  &  6.1 $\t 10^{-4}$     & 1.8 $\t 10^{-7}$  \\
\hline \hline
\end{tabular}
\end{center}\label{tab:table1}
\end{table}
We therefore conclude that, for this model, the value of $M$ lies in
the GUT range while the inflaton mass $m\ll M$ is of the order of
(though slightly less than) that in the simplest inflationary model
with $V=m^2\phi^2/2$. Note that for these values of parameters, all
inequalities (\ref{k1}), (\ref{k2}) and (\ref{k3}) are satisfied
(actually, in the case of (\ref{k2}) the quantity that should be
sufficiently large is the number of e-folds after transition
$\cal{N}$ (\ref{eq:N})).
\section{Conclusions and discussion}
The rapid advance and sophistication of cosmology experiments, most
notably those associated with measuring fluctuations in the cosmic
microwave background and which purport to throw light on the physics
of the very early universe, necessitate a close examination of
different possibilities for the generation of primordial
fluctuations responsible for the CMB signal. In this paper we have
demonstrated the possibility of a new kind of perturbation spectrum
generated during inflation. We have shown that, if during inflation,
the effective mass of the inflaton changes rapidly, then this change
results in a universal local feature being imprinted onto the
primordial spectrum of density perturbations. Namely, a sudden
change in $m_{\rm eff}^2$ satisfying the condition $|[m_{\rm
eff}^2]|\ll H^2$ leads to a break in the spectral slope --
equivalently -- to a step in the value of the primordial spectral
index $n_s$. This break is accompanied by rapid oscillations
decaying both in $P_{\cal R}(k)$ and $n_s(k)$ away from the
transition point. These oscillations are rather small in magnitude
as compared to all exact solutions for perturbation spectra with
features considered before. The amplitude of the running of the
spectral index is rather large, $\sim (n_s-1)$, at the transition
point but decays away from it, too.
The precise location of the step in $n_s$ and its amplitude are free
parameters of this model whose values must be set after comparing
the predictions of this scenario with observations. Note also we
have not specified the form of the background power spectrum
$\mathcal{P_R}_0(k)$ on which our feature is superimposed. The form
of $\mathcal{P_R}_0(k)$ must clearly be derived from a concrete
physical model. (One might even conjecture, for arguments sake, that
$\mathcal{P_R}_0(k)$ could contain additional features generated by
other physical effects such as the existence of a radiative epoch
prior to inflation, etc.)
We have also demonstrated that a field--theoretic model which can
give rise to a step in $V''(\phi)$ is similar to that used to end
inflation in the hybrid inflationary model, though different values
of its parameters are required in our case which should satisfy the
inequalities (\ref{k1}), (\ref{k2}) and (\ref{k3}). It describes a
fast second--order phase transition during inflation that occurs in
some other scalar field weakly coupled to the inflaton. Some
estimates of the values of the spectral index in this model are
given and analytical formulae relating $n_s$ to the fundamental
parameters in the inflationary potential are derived; see
(\ref{eq:n1}) \& (\ref{eq:n2}). The reader may also like to note
that while the treatment in section 3 is quite general and allows
the running of the spectral index to be positive (${\tilde\alpha} >
0$) as well as negative (${\tilde\alpha} < 0$), the microphysical
model in section 4 favours  ${\tilde\alpha} < 0$ which is suggested
by (\ref{eq:delta_n}). To conclude, the scenario discussed in this
paper suggests that the observed running of the spectral index in
the WMAP data may be caused by a fast second order phase transition
which occurred during inflation. A detailed comparison of this model
with observations remains an important problem for further study.
\section{Acknowledgements}
AS was partially supported by the Russian Foundation for Basic
Research, grant 05-02-17450, and by the Research Program
``Elementary Particles" of the Russian Academy of Sciences. He also
thanks the Inter-University Centre for Astronomy and Astrophysics,
Pune and the Yukawa Institute for Theoretical Physics, Kyoto
University, Kyoto for hospitality during the beginning and the
middle of this project correspondingly.


\begin{thebibliography}{99}
%
\bibitem{wmap}
D. N. Spergel, L. Verde, H. V. Peiris {\it et al.}, Astroph. J. Suppl.
{\bf 148}, 175 (2003) [arXiv:astro-ph/0302209]; \\
D. N. Spergel, R. Bean, O. Dor\'e {\it et al.}, Astroph. J. Suppl.
{\bf 170}, 377 (2007) [arXiv:astro-ph/0603449].
%
\bibitem{bond_boom}
C. J. MacTavish, P. A. R. Ade, J. J. Bock {\it et al.},
Astroph. J. {\bf 647}, 799 (2006) [arXiv:astro-ph/0507503].

\bibitem{archeops}
A. Benoit, P. Ade, A. Amblard {\it et al.}, Astron. Astroph. {\bf 399},
L25 (2003) [arXiv:astro-ph/0210306].

\bibitem{st05}
A. A. Starobinsky, JETP Lett. {\bf 82}, 169 (2005)
[arXiv:astro-ph/0507193].

\bibitem{pert}
V. F. Mukhanov and G. V. Chibisov, JETP Lett. {\bf 33}, 532 (1981); \\
S. W. Hawking, Phys. Lett. B {\bf 115}, 295 (1982); \\
A. A. Starobinsky, Phys. Lett. B {\bf 117}, 175 (1982); \\
A. H. Guth and S.-Y. Pi, Phys. Rev. Lett. {\bf 49}, 1110 (1982).

\bibitem{models}
A. A. Starobinsky, Phys. Lett. B {\bf 91}, 99 (1980); \\
A. D. Linde, Phys. Lett. B {\bf 108}, 389 (1982); \\
A. Albrecht and P. J. Steinhardt, Phys. Rev. Lett. {\bf 48}, 1220 (1982); \\
A. D. Linde, Phys. Lett. B {\bf 129}, 177 (1983).

\bibitem{hybrid}
A. D. Linde, Phys. Rev. D {\bf 49}, 748 (1994)
[arXiv:astro-ph/9307002].

\bibitem{PE06}
H. Peiris and R. Easther, JCAP {\bf 0610}, 017 (2006)
[arXiv:astro-ph/0609003]; \\
J. D. Barrow, A. R. Liddle and C. Pahud, Phys. Rev. D. {\bf 74},
127305 (2006) [arXiv:astro-ph/0610807].

\bibitem{EP06}
R. Easther and H. Peiris, JCAP {\bf 0609}, 010 (2006)
[arXiv:astro-ph/0604214]; \\
B. Feng, J.-Q. Xia and J. Yokoyama, arXiv:astro-ph/0608365.

\bibitem{planck} {\em The Scientific Programme of Planck},
arXiv:astro-ph/0604069.

\bibitem{KLS85}
L. A. Kofman, A. D. Linde, A. A. Starobinsky, Phys. Lett. B {\bf 157}
361 (1985); \\
J. Silk and M. S. Turner, Phys. Rev. D {\bf 35}, 419 (1987); \\
L. A. Kofman and D. Y. Pogosyan, Phys. Lett. B {\bf 214}, 508 (1988); \\
V. F. Mukhanov and M. I. Zelnikov, Phys. Lett. B {\bf 263}, 169 (1991); \\
D. Polarski and A. A. Starobinsky, Nucl. Phys. B {\bf 385}, 623 (1992); \\
D. Polarski and A. A. Starobinsky, Phys. Rev. D {\bf 50}, 6123 (1994)
[arXiv:astro-ph/9404061]; \\
J. A. Adams, G. G. Ross and S. Sarkar, Nucl. Phys. B {\bf 503}, 405
(1997) [arXiv:hep-ph/9704286].

\bibitem{KL87}
L. A. Kofman and A. D. Linde, Nucl. Phys. B {\bf 282}, 555 (1987).

\bibitem{SBB89}
D. S. Salopek, J. R. Bond and J. M. Bardeen, Phys. Rev. D {\bf 40}, 1753
(1989).

\bibitem{S92}
A. A. Starobinsky, JETP Lett. {\bf 55}, 489 (1992).

\bibitem{S85}
A. A. Starobinsky, JETP Lett. {\bf 42}, 152 (1985).

\bibitem{YY06}
M. Yamaguchi and J. Yokoyama, Phys. Rev. D {\bf 74}, 043523 (2006)
[arXiv:hep-ph/0512318].

\bibitem{S98}
A. A. Starobinsky,  "\emph{Large Scale Structure: Tracks and
Traces}", in: Proc. of the 12th Potsdam Cosmology Workshop (15-20
Sept. 1997), eds. V. Muller, S. Gottlober, J. P. Mucket, J.
Wambsganss (Singapore: World Scientific), 1998, pp. 375-380
[arXiv:astro-ph/9808152].

\bibitem{G05}
J.-O. Gong, JCAP {\bf 0507}, 015 (2005) [arXiv:astro-ph/0504383].

\bibitem{PNN94}
P. Ivanov, P. Naselsky and I. Novikov, Phys. Rev. D {\bf 50}, 7173 (1994).

\bibitem{GLW96}
J. Garc\'{i}a-Bellido, A.D. Linde and D. Wands, Phys. Rev. D {\bf
54}, 6040 (1996) [arXiv:astro-ph/9605094].

\bibitem{step_pot}
J. A. Adams, B. Cresswell and R. Easther, Phys. Rev. D {\bf 64},
123514 (2001) [arXiv:astro-ph/0102236]; \\
H. V. Peiris, E. Komatsu, L. Verde \etal, Astrophys. J. Suppl. {\bf
148}, 213 (2003) [arXiv:astro-ph/0302225].

\bibitem{HCMS07}
J. Hamann, L. Covi, A. Melchiorri and A. Slosar, Phys. Rev. D {\bf
76}, 023503 (2007)  [arXiv:astro-ph/0701380].

\bibitem{LPS98}
J. Lesgourgues, D. Polarski and A. A. Starobinsky, MNRAS {\bf 297},
769 (1998) [arXiv:astro-ph/9711139].

\bibitem{step_deriv}
A. Shafieloo and T. Souradeep, Phys.Rev. D {\bf 70} (2004) 043523
[arXiv:astro-ph/0312174]; \\
R. Sinha and T. Souradeep, Phys.Rev. D {\bf 74} (2006) 043518
[arXiv:astro-ph/0511808]; \\
A. Shafieloo, T. Souradeep, P. Manimaran {\it et al.}, Phys. Rev. D
{\bf 75}, 123502 (2007) [arXiv:astro-ph/0611352]; \\
A. Shafieloo and T. Souradeep, arXiv:0709.1944 [astro-ph].

\bibitem{BBKP03}
D. Blais, T. Bringmann, C. Kiefer and D. Polarski, Phys. Rev. D {\bf
67} (2003) 024024 [arXiv:astro-ph/0206262].

\bibitem{SL93}
E.D. Stewart and D.H. Lyth, Phys. Lett. B {\bf 302}, 171 (1993)
[arXiv:gr-qc/9302019].


\bibitem{subir}
P. Hunt and S. Sarkar, Phys. Rev. D {\bf 70}, 103518 (2004)
[arXiv:astro-ph/0408138]; Phys. Rev. D {\bf 76}, 123504 (2007)
[arXiv:0706.2443 [astro-ph]].
\bibitem{sasaki83}
M. Sasaki, Progr. Theor. Phys. {\bf 70}, 394 (1983).


\bibitem{LV07}
J. Lesgourgues and W. Valkenburg, Phys. Rev. D {\bf 75}, 123519
(2007) [arXiv:astro-ph/0703625]; \\
 J. Lesgourgues, A. A. Starobinsky and  W. Valkenburg,
J. Cosmol. Astropart. Phys. 01 (2008) 010 [arXiv:0710.1630
[astro-ph]].

\bibitem{mukhanov88}
V.F. Mukhanov, Sov. Phys. - JETP {\bf 67}, 1297 (1988).

\bibitem{lifshitz}
E.M. Lifshitz, Zh. Eksp. Teor. Fiz. {\bf 16}, 587 (1946); L.D.
Landau and E.M. Lifshitz, {\em The classical theory of fields},
Pergamon, New York, 1976.m

\bibitem{carroll92}
S.M. Carroll, W.H. Press and E.L. Turner, Ann. Rev. Astron. Astrophys. {\bf 30}, 499 (1992).

\bibitem{bunn96}
E.F. Bunn, A.R. Liddle and M. White, Phys.Rev. D {\bf 54}, R5917
(1996) [arXiv:astro-ph/9607038].

\bibitem{ll00}
A.R. Liddle and D.H. Lyth, \emph{Cosmological Inflation and Large
Scale Structure}, Cambridge University Press (2000).


\end{thebibliography}
\end{document}